\documentclass{iau307}
\usepackage{graphicx}
\usepackage{natbib}
\usepackage{url}
\usepackage{dtklogos}
\bibpunct{(}{)}{;}{a}{}{,}

\title[A New Class of WRs] 
{A New Class of Wolf-Rayet Stars: WN3/O3s}

\author[Massey, Neugent, Morrell, \& Hillier]   
{Philip Massey$^1$,
 Kathryn F. Neugent$^1$,
 Nidia Morrell$^2$,
 \and D.~John~Hillier$^3$}

\affiliation{$^1$Lowell Observatory \\ email: {\tt phil.massey@lowell.edu} \\[\affilskip]
$^2$Las Campanas Observatory, Carnegie Observatories
$^3$Department of Physics and Astronomy \& Pittsburgh Particle Physics, Astrophysics, and Cosmology Center, University of Pittsburgh}

\pubyear{2014}
\volume{307} 
\pagerange{}
\jname{New windows on massive stars: asteroseismology, interferometry, and spectropolarimetry}
\editors{G. Meynet, C. Georgy, J.H. Groh \& Ph. Stee, eds.}

\begin{document}

\maketitle

\begin{abstract}
Our new survey for Wolf-Rayet stars in the Magellanic Clouds is only 15\% complete but has already found 9 new WRs in the LMC.  This suggests
that the total WR population in the LMC may be underestimated by 10-40\%.  Eight of the nine are WNs, demonstrating that the ``observed" WC to WN ratio is too large, and is biased towards WC stars.  The ninth is another rare WO star, the second we have found in the LMC in the past two years. Five (and possibly six) of the 8 WNs are of a new class of WRs, which pose a significant challenge to our understanding.  Naively we would classify these stars as ``WN3+O3V," but there are several reasons why such a pairing is unlikely, not the least of which is that the absolute visual magnitudes of these stars are {\it faint}, with $M_V\sim -2.3$ to $-3.1$.  We have performed a preliminary analysis with CMFGEN, and we find that (despite the faint visual magnitudes) the bolometric luminosities of these stars are normal for early-type WNs.  Our fitting suggests that these stars are evolved, with significantly enriched N and He.  Their effective temperatures are also normal for early-type WNs.   What is unusual about these stars is that they have a surprisingly small mass-loss rate compared to other early-type WNs.  How these stars got to be the way they are (single star evolution?  binary evolution?) remains an open question.  For now, we are designating this class as WN3/O3, in analogy to the late-type WN ``slash" stars.
\end{abstract}

\firstsection 
\section{A Modern Survey for Wolf-Rayet Stars in the Magellanic Clouds}

One of the long-standing tests of massive star evolutionary models is how well they do in predicting the observed ratios of various types of evolved massive stars as a function of metallicity \citep[see, e.g.,][]{Maeder80, MJ98, MM05, Meynet07, Eldridge08, M31WRs}.  Of these tests, possibly the most robust is that of comparing the relative number of WC- and WN-type WRs.  Observationally, this quantity has long been known to be low in the SMC, higher in the LMC, and higher still in the Milky Way, in accord with the progression of metallicity.  \citet{VC80} were  the first to attribute this to the effects of metallicity on the mass-loss rates of main-sequence massive stars.
However, as shown in many previous studies \citep[e.g.,][]{MC83, AM85, MJ98, M33WRs, M31WRs}, the ``observed" ratios of WCs to WNs can
readily be biased towards stars of WC type, as the strongest lines in WCs are significantly ($\sim$10$\times$) stronger than the strongest lines
in WNs \citep{MJ98}.  We have therefore begun new surveys for WRs in the star-forming galaxies of the Local Group  (i.e., M33, \citeauthor{M33WRs} \citeyear{M33WRs}, and
M31, \citeauthor{M31WRs} \citeyear{M31WRs}) to allow more meaningful comparisons with the new generation of evolutionary
models now becoming available \citep[e.g.,][]{Eldridge08,Sylvia12,Cyril13}.  

As part of this process, our attention was drawn to the fact that the Geneva rotating models at LMC metallicity predict a  WC to WN ratio of 0.09 \citep{M31WRs} while the observed ratio is 0.23 according to the \citet{BAT99} catalog (BAT99), as updated by \citet{WO}.  Is this a problem with the
single-star Geneva models, or could the problem be observational? For many years, our knowledge of the WR population of the Magellanic Clouds has been considered essentially complete. We realized that in the 15 years since the BAT99 catalog was published, there were seven new WRs found in the LMC, 6 of them of were of WN type.  The other one was a rare WO star, found by ourselves \citep{WO}.  The WO star has very strong lines, and was found as part of a spectroscopic study of the stellar content of Lucke-Hodge 41, the home of S Doradus and many other massive stars.  This discovery served as pretty much the last straw: we felt it behooved us to conduct a modern search for Wolf-Rayet stars in the Magellanic Clouds \citep{MCWRs}.

It was clear from the onset {\it how} we needed to do this: we would use the same sort of interference filters we had used so successfully to survey M33 and M31 for WRs \citep{M33WRs, M31WRs}.  We even knew {\it where} we wanted to do this: the Swope 1-m telescope had both a good image scale and a large FOV.  Nevertheless, it would take about 800 fields and several hundred hours of observing to completely cover both Clouds.  Fortunately they have just replaced the camera with one with an even larger FOV and much reduced overhead.  Nevertheless this is a long term, multi-year project.  In our first year (with 6 excellent nights) we were able to survey about $\sim$15\% of each Cloud.  Image-subtraction techniques then allowed us to identify stars which
were brighter in a C~III $\lambda 4650$ or a He II $\lambda 4686$ filter relative to neighboring continuum.  We had concentrated on fields where WR stars were already known (following the philosophy attributed apocryphally to ``Slick" Willie Sutton) and readily recovered all of the previously known WRs, except in the most crowded regions of the R136 cluster.  We also re-discovered many previously known planetary nebulae and Of-type stars, both of which
would have He II $\lambda 4686$ in emission.  And, much to our relief, we also found a number of previously unknown WR candidates. Follow-up spectroscopy with Magellan allowed us to confirm 9 new WRs in the LMC in our first observing season.  This increases the number of known WRs in the LMC by about 6\%, suggesting that the total WR population of the LMC has been underestimated by 10-40\%.  While impressive, the greater significance is not in the quantity but the quality of what we found.

\section{What We Found}
Of the 9 newly found WRs in the LMC, 8 of these are WN stars.  The other is another WO star.  The WO is located only 9$^{\prime\prime}$ from the
one we found two years ago by accident in Lucke-Hodge 41 \citep{WO}.    Of the 8 WNs, 2 are normal WN3+mid-to-late O-type binaries; we were
able to obtain two epochs of radial velocities that show the absorption and emission moving in anti-phase, as one would expect.

Five others (and possibly the sixth) are, however, quite peculiar.  Naively we would classify their spectra as WN3+O3V.  However, such a pairing
would be unlikely for several reasons, the most unarguable one being that the absolute visual magnitudes of these six stars are quite faint, with $M_V\sim -2.3$ to $-3.1$.  This is much fainter than that of a typical O3~V star, with $M_V\sim -5.4$ \citep{Conti88}.  Furthermore, we had two
radial velocity epochs for three of these which failed to show variations.  We will therefore refer to these stars as WN3/O3 WRs, with the slash
reminiscent of the ``transition" Ofpe/WN9 stars \citep{BW89,BB90}.

Could both the absorption and emission be coming from  a single object?  To explore this, we attempted to fit our optical spectrum of one of these stars (LMC170-2) with CMFGEN \citep{CMFGEN}, a stellar atmosphere code designed for hot stars with stellar winds where the usual assumptions of plane-parallel geometry and local thermodynamic equilibrium no longer hold. We found that we could obtain a very good fit with a single set of parameters as given in Table~1.  Portions of the fit are shown in Fig.~1.  A slightly different model also produced an adequate fit, and we give these parameters in Table 1 as well, to demonstrate the current uncertainties.

\begin{figure}
\begin{center}
\includegraphics[width=0.8\textwidth]{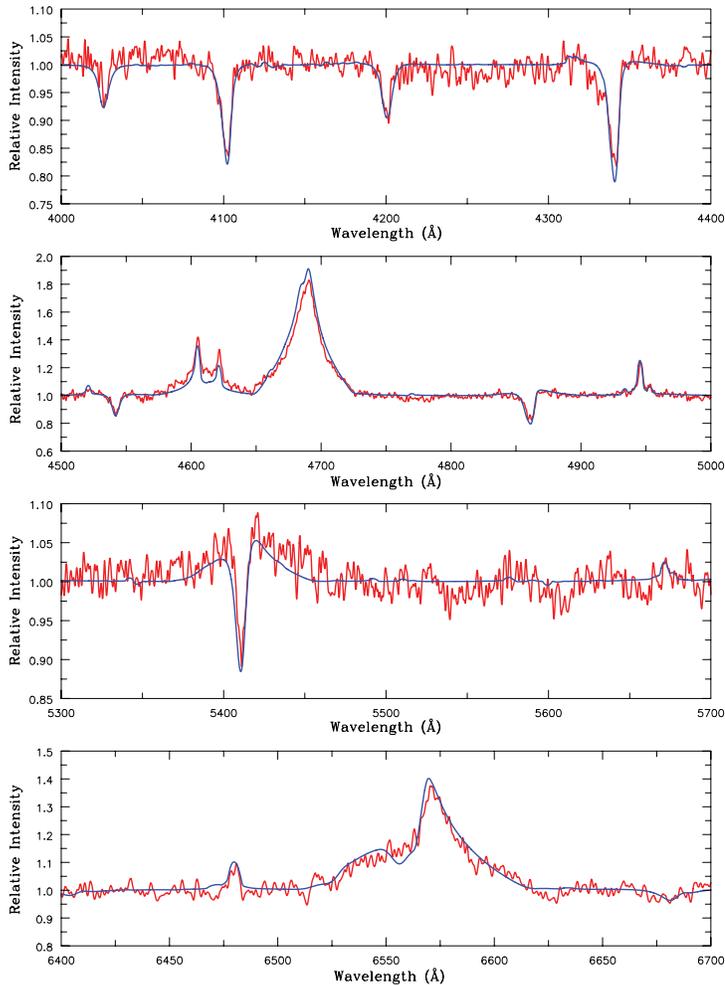} 
\caption{Our best CMFGEN fit to the optical spectrum of LMC170-2   From \citet{MCWRs} and used by permission.}
\label{fig1}
\end{center}
\end{figure}

\begin{table}
\begin{center}
\caption{Parameters of our CMFGEN fits.}
\label{tab1}
\begin{tabular}{l c c}\hline 
\textbf{Parameter} & \textbf{Best Fit} & \textbf{Pretty Good Fit} \\ 
\hline
Teff (K) & 100,000 & 80,000 \\
$L/L_\odot$ & $4\times10^5$& $2.0\times10^5$\\
$\dot{M}^1$ ($M_\odot$/yr) & $1.2\times10^{-6}$& $7.6\times10^{-7}$ \\
He/H (by \#)  & 1.0 & 0.5 \\
N &$10.0\times$ solar & $5.0\times$ solar \\
C, O & $0.05\times$ solar & $0.05\times$ solar\\
\hline
\end{tabular}
\end{center}
\vspace{1mm}
\scriptsize{
{\it Notes:}\\
$^1$Assumes a clumping filling factor of 0.1, $v_\infty$=2400 km/sec, and $\beta$=0.8. \\
}
\end{table}

\begin{figure}
\begin{center}
\includegraphics[width=0.35\textwidth]{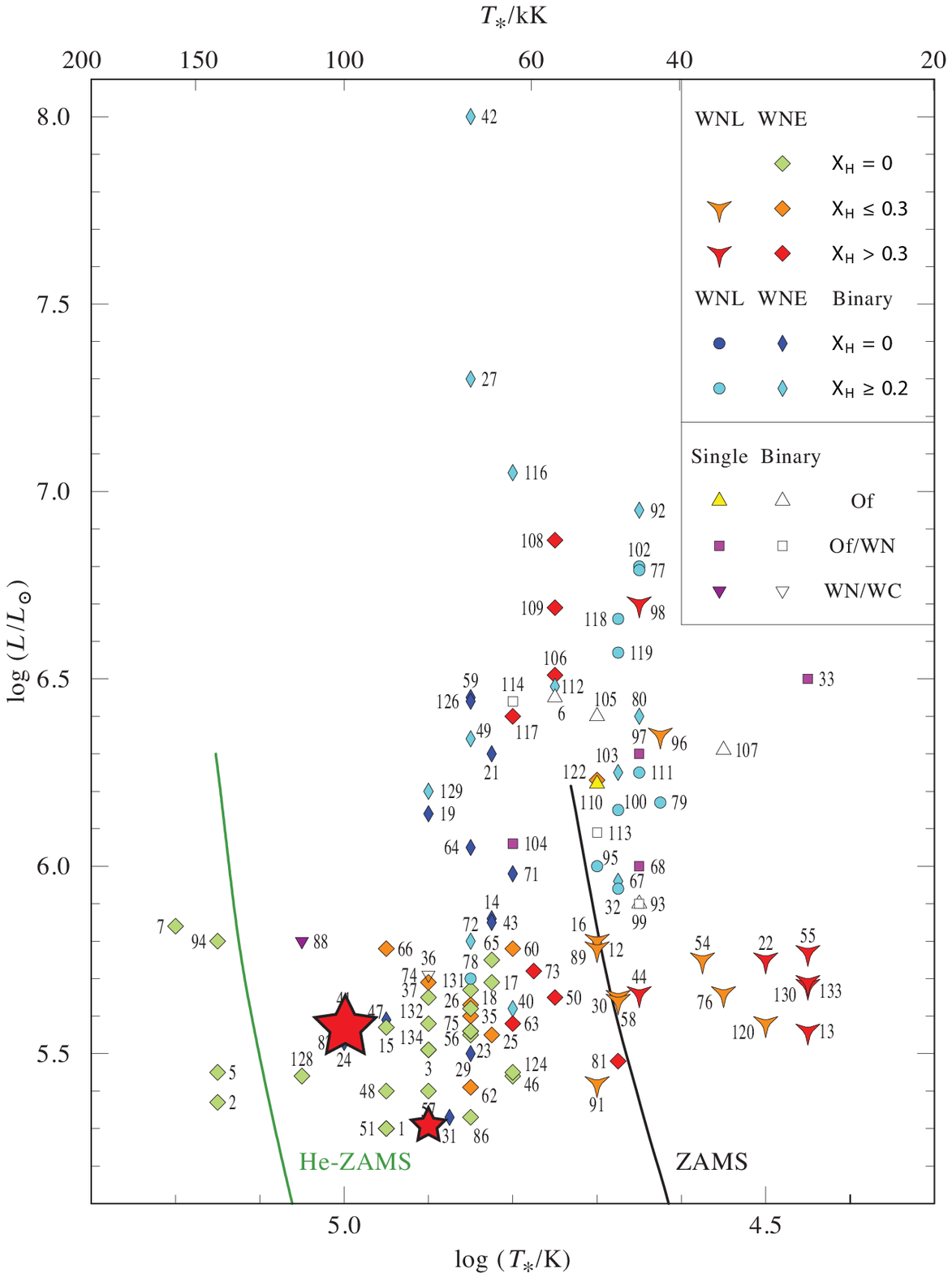} 
\includegraphics[width=0.4\textwidth]{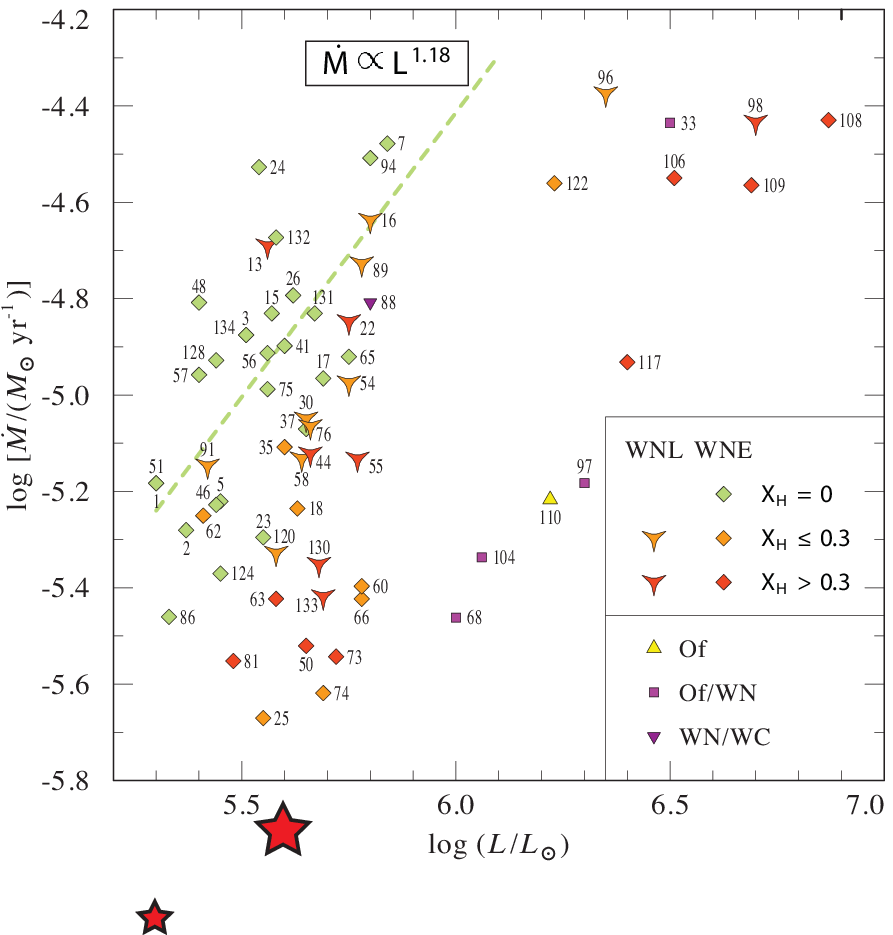} 
\caption{How the parameters of our fit of the WN3/O3 star LMC170-2 compare to those of other LMC WNs analyzed by \citet{Potsdam}.  Our best fit is shown by a large star, while our ``pretty good" fit is shown by a smaller star.  Adapted from Figs.\ 6 and 7 of \citet{Potsdam} and used by permission.}
\label{fig2}
\end{center}
\end{figure}

In Fig.~2 we show how these parameters compare to that of the LMC WN stars recently analyzed by \citet{Potsdam}.  What we find is that the
effective temperature and luminosity are normal for early-type WN stars (Fig.~2, left), but that the mass-loss rates are significantly lower (Fig.~2, right).  In many ways this makes sense, as it suggests that the winds are weak enough that we see something like a normal stellar photosphere.

Of course, this now raises significant new questions:  Why are the mass-loss rates low, and how did these stars evolve? Are binary models needed to produce such objects, or are these the hitherto unrecognized products of single-star evolution?   If the former, then where is the spectroscopic signature of the companion?
 
 We note that at this point the physical parameters are not well constrained (compare the two models in Table 1): the optical spectra lack  multiple ionization stages of the same species, impacting our ability to constrain the effective temperature.  We just learned that we were successful in obtaining time on {\it HST} to observe these objects in the UV; these spectra will provide data on additional ionization states (such as N~IV $\lambda 1718$) as well as provide key diagnostics of the stellar wind, such as the C~IV $\lambda 1550$ resonance line.  We also plan to obtain higher S/N optical data with Magellan, which should also help.  We will continue to monitor these stars for radial velocity variations. We will be carrying out our survey for WRs in the rest of the Magellanic Clouds, and this will show us to see how many more of these WN3/O3s are out there.  
 
 And, of course, we do not know what other surprises await!  In addition to these new WRs our study has also revealed two O8f?p stars, further examples
 of this rare class of  magnetically-braked oblique rotators \citep{Nolan2010}. We have ten further nights scheduled on the Swope in the next Magellanic Cloud observing season (late 2014) with time for follow-up spectroscopy.  So, stayed tuned!
 
 This work has been supported by the National Science Foundation under AST-1008020 and by Lowell Observatory's research support fund, thanks to generous donations by Mr.\ Michael Beckage and Mr.\ Donald Trantow.  D.J.H. acknowledges support from STScI theory grant HST-AR-12640.01.  We appreciate the fine support we always receive at Las Campanas Observatory, where these observations were carried out.
 We are also grateful to our hosts here in Geneva for providing an opportunity to discuss this work in such a nice setting!

\bibliographystyle{iau307}
\bibliography{MyBiblio}

\begin{discussion}

\discuss{Sundqvist}{It seems to me that another interpretation of your two very interesting objects would be that they are just slightly evolved, still hydrogen-burning O-star with unusual high $T_\mathrm{eff}$. Is this something you have considered?}

\discuss{Massey}{It was our first thought -- that maybe these were not evolved objects. However, our modelling has shown that nitrogen is strongly enhanced, and C and O are way down. The He/H ratio is about 1 by number. I can't say what's going on in the core. But we have never seen an O star like this, and naively to me the $\mathrm{N}_\mathrm{V}$ and $\mathrm{He}_\mathrm{II}$ emission line strengths are similar to those of WN3 stars.}

\discuss{Najarro}{Phil, were you able to get a handle on $\log(g)$ (masses) for the objects?}

\discuss{Massey}{We adopted a $\log(g)$ of $5.0$.  That would lead to a mass of 15$M_\odot$, but the uncertainties are of order factors of 2 or 3. John felt he could rule out a $\log(g)$ of $5.5$ for a temperature of $100000\,\mathrm{K}$, but it isn't well constrained.  We hope with better optical data (higher S/N) and UV data with {\it HST} we can get more solid physical parameters.} 

\end{discussion}

\end{document}